# Um Desenvolvimento Numérico nas Equações Dinâmicas de Sólitons em Fibras Óticas

# A Numerical Development in the Dynamical Equations of Solitons into Optical Fibers

Diogo Albino de Queiroz[1]; Paulo Laerte Natti[2], Neyva Maria Lopes Romeiro[3]; Érica Regina Takano Natti[4]

## Resumo

Desenvolvemos e avaliamos um procedimento numérico para um sistema de equações diferenciais não-lineares, que descrevem a propagação de sólitons em fibras óticas dielétricas ideais. Este problema tem soluções analíticas conhecidas. A resolução numérica do sistema é implementada por meio do método de elementos finitos, utilizando métodos de estabilização do tipo *Streamline Upwind Petrov-Galerkin* (SUPG) e *Consistent Approximate Upwind* (CAU). Comparando as soluções analíticas e numéricas, verificou-se que o procedimento numérico descreve adequadamente a dinâmica deste sistema.

Palavras-chave: Fibra ótica, sóliton, método de elementos finitos, *Streamline Upwind Petrov-Galerkin*, *Consistent Approximate Upwind*.

## Abstract

We develop and evaluate a numerical procedure for a system of nonlinear differential equations, which describe the propagation of solitons in ideal dielectric optical fibers. This problem has analytical solutions known. The numerical solutions of the system is implemented by the finite element method, using methods of stabilization such as Streamline Upwind Petrov-Galerkin (SUPG) and Consistent Approximate Upwind (CAU). Comparing the numerical and analytical solutions, it was found that the numerical procedure adequately describes the dynamics of this system.

[1] Bolsista de Iniciação Científica IC/UEL. Aluno de graduação em Ciência da Computação do Departamento de Computação da Universidade Estadual de Londrina, Londrina – PR – Brasil. Email: daqueiroz@dc.uel.br.
[2] Professor Doutor do Departamento de Matemática da Universidade Estadual de Londrina, Londrina – PR – Brasil. Email: plnatti@uel.br.
[3] Professora Doutora do Departamento de Matemática da Universidade Estadual de Londrina, Londrina – PR – Brasil. Email: nromeiro@uel.br.
[4] Professora Doutora da Pontifícia Universidade Católica do Paraná, Londrina – PR – Brasil. Email: erica.takano@uol.com.br.

# Introdução

Recentemente, considerável atenção tem sido dada ao estudo de uma classe de equações diferenciais de campos acoplados com não-linearidades de segunda ordem nos campos. Uma grande variedade de fenômenos pode ser descrita por esses sistemas de equações, em particular, fenômenos não-lineares em ótica. Comunicação ótica (HAUS; WONG, 1996), *light bullets* (MALOMED et al., 2005), aparelhos óticos tais como interferômetros do tipo Mach-Zehnder (ASSANTO et al., 1993), *couplers* direcionais (LEDERER et al., 1990) e outros (TAYLOR, 1992, HASEGAWA; KODAMA, 1995) são descritos por tais equações.

Nosso interesse em estudar esses sistemas de equações é aplicá-las ao estudo da estabilidade de sinais em fibras óticas com geometria transversal retangular, nas quais as ondas estacionárias do tipo sólitons são possíveis (GALLEAS et al., 2003, YMAI et al., 2004).

Os sólitons podem ser utilizados em comunicações a longa distância sem a necessidade de estações repetidoras, inclusive em comunicações transoceânicas, como foi proposto por Hasegawa (1984). A comunicação convencional utiliza sinais lineares, no formato NRZ (*nonreturn to zero*), que são regenerados eletronicamente, e, então, retransmitidos. A localização e a distância das estações repetidoras são escolhidas, de modo que ainda seja possível recuperar o sinal com uma taxa de erro menor que 1 bit em $10^9$ bits. Atualmente, essa distância varia entre 120-150 Km. Por outro lado, Nakazawa e colaboradores (1991) haviam propagado sólitons em laboratório, num sistema fechado (em *loop*) de fibras, a distâncias superiores a $10^6$ Km já em 1991, utilizando moduladores de amplitude.

As propriedades dielétricas que fibras óticas devem apresentar, de modo que a comunicação via sólitons seja mais estável e eficaz que a comunicação via sinais lineares no formato NRZ, é de grande interesse comercial e tecnológico. Por apresentar diversas aplicações potenciais em Telecomunicações, estudos sobre a propagação e a estabilidade de sólitons têm tido um grande crescimento nos últimos quinze anos (HAUS; WONG, 1996, QIN; DAI; ZHANG, 2005).

Neste trabalho faremos um estudo numérico de sistemas de equações diferenciais, assintoticamente equivalentes à equação de Schrödinger não-linear (ZAKHAROV; SABAT, 1972, HASEGAWA; TAPPERT, 1973, MENYUK; SCHIEK; TORNER, 1994). Essas equações descrevem a evolução de ondas tipo sólitons em fibras óticas com não-linearidades quadráticas, também chamadas de guias dielétricos do tipo $\chi^{(2)}$. Este estudo será implementado para um sistema simplificado de equações diferenciais parciais (EDP's), o qual apresenta soluções analíticas conhecidas (GALLEAS et al., 2003). Desse modo iremos ajustar e avaliar um procedimento numérico desenvolvido por Romeiro e colaboradores, baseado no método de elementos finitos com formulações do tipo *Streamline Upwind Petrov-Galerkin* - SUPG (GALEÃO; CARMO, 1988, BROOKS; HUGHES, 1982, HUGHES; FRANCA; MALLET, 1986) e do tipo *Consistent Approximate Upwind* - CAU (ALMEIDA; GALEÃO, 1993, GALEÃO et al., 2004).

A partir deste trabalho, será possível estudar a solução numérica de sistemas de EDP's mais complexos, que descrevem de forma realista a propagação de sólitons em materiais dielétricos. Esses sistemas, ditos realistas, não têm soluções analíticas conhecidas, de forma que o tratamento numérico é uma maneira possível para simular e otimizar a propagação destas ondas em função das propriedades dielétricas dos materiais (YMAI et al., 2004).

Para o desenvolvimento deste trabalho propomos três etapas. Na primeira etapa, a partir de um sistema de EDP's, que descreve a propagação de sóliton ideais em fibras óticas do tipo $\chi^{(2)}$, deduzimos um sistema de equações diferenciais ordinárias (EDO's) equivalente. Na segunda etapa, apresentamos as diretrizes, referentes ao procedimento numérico, para a resolução do sistema de EDO's obtido. Na terceira etapa, comparando graficamente as soluções analíticas conhecidas com as soluções numéricas obtidas, analisamos o método numérico quando aplicado ao sistema de equações diferenciais discutidos neste trabalho.

# Dinâmica Longitudinal de Campos Acoplados em Dielétricos $\chi^{(2)}$

O sistema de EDP's não-lineares, obtido a partir das equações de Maxwell, descreve a evolução longitudinal de duas ondas eletromagnéticas acopladas (modos fundamental e segundo harmônico) em guias de ondas (fibras óticas) com não-linearidades do tipo $\chi^{(2)}$. Esse sistema é apresentado nas equações (1) e (2) (GALLEAS et al., 2003)

$$i\frac{\partial a_1}{\partial \xi} - \frac{r}{2}\frac{\partial^2 a_1}{\partial s^2} + a_1^* a_2 \exp(-i\beta\xi) = 0 \quad (1)$$

$$i\frac{\partial a_2}{\partial \xi} - i\delta\frac{\partial a_2}{\partial s} - \frac{\alpha}{2}\frac{\partial^2 a_2}{\partial s^2} + a_1^2 \exp(i\beta\xi) = 0 \quad (2)$$

onde $a_1(s,\xi)$ e $a_2(s,\xi)$ são variáveis complexas e representam as amplitudes normalizadas das ondas fundamental e segundo harmônico, respectivamente.

O sistema dado em (1) e (2) apresenta soluções analíticas do tipo sólitons para certas faixas de valores de $r$, $\alpha$, $\beta$ e $\delta$ (YMAI et al., 2004). Vejamos como estas quantidades caracterizam as propriedades não-lineares (dielétricas) da fibra ótica. No limite $|\beta| \to \infty$, as equações diferenciais (1) e (2) reduzem-se à equação de Schrödinger não-linear (NLSE) (MENYUK; SCHIEK; TORNER, 1994), correspondendo ao limite de incompatibilidade de fase (ondas fora de fase). Consequentemente, a quantidade $\beta$ mede a taxa de geração do segundo harmônico, ou ainda, a intensidade de não-linearidade do material. A quantidade $\alpha$ mede a dispersão relativa das velocidades de grupo das ondas fundamental e segundo harmônico no material (fibra ótica). Para valores de $|\alpha| > 1$, a onda do segundo harmônico possui dispersão maior que a onda fundamental e para valores de $|\alpha| < 1$ é a onda fundamental que tem dispersão maior. A quantidade $r$ é um indicador do regime de dispersão da onda fundamental. Quando temos $r = +1$, a onda fundamental encontra-se no regime de dispersão dito normal, porém se $r = -1$, a onda fundamental encontra-se no regime de dispersão dito anômalo. Enfim, a quantidade $\delta$ mede a diferença das velocidades de grupo dos modos fundamental e segundo harmônico, e assume valores não-nulos, por exemplo, em meios anisotrópicos, quando os vetores de Poynting destes modos estão desalinhados (*walk-off wave*). No trabalho de Galleas e colaboradores (2003), encontra-se uma descrição pormenorizada da dedução da solução das equações (1) e (2) e da interpretação das quantidades dielétricas $\alpha$, $\beta$, $\delta$ e $r$, relacionando-as com as propriedades óticas da fibra.

Observamos que podemos escolher as características da onda sóliton a ser propagada na fibra ótica (velocidade, largura, amplitude, estabilidade, etc..), selecionando ou propondo materiais com as propriedades dielétricas $\alpha$, $\beta$, $\delta$ e $r$ adequadas.

As soluções analíticas tipo sólitons das equações (1) e (2) são conhecidas (MENYUK; SCHIEK; TORNER, 1994, GALLEAS et al., 2004) e apresentadas em (3) e (4),

$$a_1(s,\xi) = |\beta|^{1/2} \hat{a}_1(s,\xi) \quad (3)$$

$$a_2(s,\xi) = \hat{a}_2(s,\xi)\exp(i\beta\xi) \quad (4)$$

onde

$$\hat{a}_1 = \pm\frac{3}{2(\alpha-2r)}\sqrt{\frac{\alpha r}{|\beta|}}\left(\frac{\delta^2}{2\alpha-r}+\beta\right) \times$$

$$\text{sch}^2\left[\pm\sqrt{\frac{1}{2(2r-\alpha)}\left(\frac{\delta^2}{2\alpha-r}+\beta\right)}\left(s-\frac{r\delta}{2\alpha-r}\xi\right)\right] \times$$

$$\exp\left\{i\left[\frac{r\delta^2(4r-5\alpha)}{2(2\alpha-r)^2(2r-\alpha)}-\frac{r\beta}{2r-\alpha}\right]\xi - \frac{i\delta}{2\alpha-r}s\right\}$$

(5)

$$\hat{a}_2 = \frac{3r}{2(\alpha-2r)}\left[\frac{\delta^2}{2\alpha-r}+\beta\right] \times$$

$$\text{sch}^2\left[\pm\sqrt{\frac{1}{2(2r-\alpha)}\left(\frac{\delta^2}{2\alpha-r}+\beta\right)}\left(s-\frac{r\delta}{2\alpha-r}\xi\right)\right] \times$$

$$\exp\left\{2i\left[\frac{r\delta^2(4r-5\alpha)}{2(2\alpha-r)^2(2r-\alpha)}-\frac{r\beta}{2r-\alpha}\right]\xi - \frac{2i\delta}{2\alpha-r}s\right\}$$

(6)

Por simplicidade, consideraremos $\delta = 0$ no desenvolvimento numérico de (1) e (2). No caso da propagação de sólitons em fibras óticas ordinárias, e em situações não-críticas, o fenômeno *walk-off wave* pode ser desconsiderado (ARTIGAS; TORNER; AKHMEDIEV, 1999), o que justifica tomar $\delta = 0$ nestas situações. Assim, o sistema de EDP's torna-se separável e as soluções estacionárias de (1) e (2) têm a forma

$$a_\nu = (-1)^\nu U_\nu(s) \exp(i\kappa_\nu \xi) \quad (7)$$

onde $\nu = 1, 2$. O cálculo das derivadas parciais é dado por

$$\frac{\partial a_\nu}{\partial \xi} = (-1)^\nu i\kappa_\nu U_\nu(s) \exp(i\kappa_\nu \xi) \quad (8)$$

$$\frac{\partial a_v}{\partial s} = (-1)^v \exp(i\kappa_v \xi)\frac{\partial U_v}{\partial s} \quad (9)$$

$$\frac{\partial^2 a_v}{\partial s^2} = (-1)^v \exp(i\kappa_v \xi)\frac{\partial^2 U_v}{\partial s^2} \quad (10)$$

Cabe observar que $a_1^* = -U_1(s)\exp(-i\kappa_1\xi)$. Substituindo (7) e (10) em (1), obtém-se:

$$\kappa_1 U_1 \exp(i\kappa_1\xi) + \frac{r}{2}\exp(i\kappa_1\xi)\frac{\partial^2 U_1}{\partial s^2} -$$
$$U_1 U_2 \exp[i(-\kappa_1 + \kappa_2 - \beta)\xi] = 0 \quad (11)$$

Analogamente, substituindo (7) e (10) em (2), segue que

$$\kappa_2 U_2 \exp(i\kappa_2\xi) + \frac{\alpha}{2}\exp(i\kappa_2\xi)\frac{\partial^2 U_2}{\partial s^2} -$$
$$U_1^2 \exp[i(2\kappa_1 + \beta)\xi] = 0 \quad (12)$$

Tomando $\kappa_2 = 2\kappa_1 + \beta$, condição necessária para que (1) e (2) tenham soluções do tipo (7), obtemos o sistema de EDO's que descreve a evolução de $U_1(s)$ e $U_2(s)$, ou seja,

$$\frac{r}{2}\frac{\partial^2 U_1}{\partial s^2} + \kappa_1 U_1 - U_1 U_2 = 0 \quad (13)$$

$$\frac{\alpha}{2}\frac{\partial^2 U_2}{\partial s^2} + \kappa_2 U_2 - U_1^2 = 0 \quad (14)$$

onde todas as variáveis e constantes são reais. As constantes $r$, $\alpha$ e $\beta$ são obtidas a partir das propriedades materiais da fibra ótica, enquanto o chamado deslocamento não-linear do número de onda $\kappa_v$, para $v = 1,2$, parametriza uma família de soluções. Na próxima etapa, iremos descrever o procedimento numérico utilizado nas equações (13) e (14).

## Procedimento Numérico

Para resolver numericamente o sistema de equações diferenciais dado em (13) e (14), utilizamos o método de elementos finitos obtido a partir da formulação de Petrov-Galerkin (GALEÃO; CARMO, 1988). Sabe-se que simulações numéricas calculadas por elementos finitos usuais, tais como o método de *Galerkin*, apresentam dificuldades numéricas relacionadas à deficiência na estabilidade, gerando assim oscilações espúrias (ALMEIDA, et al., 2004). Para contornar este problema, são usados métodos estabilizados do tipo *Streamline Upwind Petrov-Galerkin* - SUPG (GALEÃO; CARMO, 1988, BROOKS; HUGHES, 1982, HUGHES; FRANCA; MALLET, 1986) e do tipo *Consistent Approximate Upwind* - CAU (ALMEIDA; GALEÃO, 1993, GALEÃO et al., 2004).

Um código computacional em FORTRAN 90, desenvolvido por um dos autores (ROMEIRO, CASTRO, LANDAU, 2003), foi ajustado para simular nosso problema. Na etapa seguinte, comparando graficamente as soluções numéricas obtidas com as soluções analíticas conhecidas, poderemos tirar conclusões sobre a metodologia numérica empregada.

## Resultados Numéricos e Analíticos

Para o código que calcula a solução numérica das equações (13) e (14), são fornecidos valores para as quantidades dielétricas $\alpha$, $\beta$ e $r$, que definem as características do sóliton a ser propagado na fibra ótica. Experimentalmente, para fibras óticas comerciais, estas quantidades podem assumir os seguintes valores: $\alpha = -1/2$, $\beta = -1/2$ e $r = -1$. É importante observar que, nas equações (13) e (14), todas as variáveis e quantidades dielétricas são adimensionais (GALLEAS et al., 2003). A variável temporal $\xi$ é fixada em $\xi = 10$, de modo que a variável espacial $s$ descreverá a forma do envelope da onda sóliton, ao longo da fibra ótica, para este "instante" fixado. Resta ainda fornecer as condições iniciais, ou ainda, o formato inicial do envelope da onda que será injetado na fibra ótica. Novamente, com o objetivo de simplificar os procedimentos, quando da comparação das soluções numérica e analítica, tomamos como condição inicial no contorno esquerdo da fibra ótica, uma onda tipo sóliton. Tomando $s = 0$ nas soluções analíticas (5) e (6), tem-se de (7) que as condições no contorno esquerdo são $U_1(0,\xi) = \frac{-3\beta\sqrt{\alpha r}}{2(\alpha - 2r)}$ e $U_2(0,\xi) = \frac{3\beta r}{2(\alpha - 2r)}$.

Para o contorno direito, o gradiente de ambas as ondas são fixados em zero, pois para distâncias $s$ suficientemente grandes as amplitudes $U_1$ e $U_2$ tendem a zero, de modo que a taxa de variação (gradiente) destas no contorno direito também tendem a zero.

Das equações (13) e (14), considerando uma partição com 30 elementos, consequentemente 31 nós, e com intervalos $\Delta s = 0.5$, obtivemos o resultado numérico que pode ser observado na figura 1. Como esperado, as condições no contorno esquerdo, no instante $\xi = 10$, fazem com que a onda sóliton esteja centrada em $s = 0$ e tenda a zero para valores maiores da variável $s$.

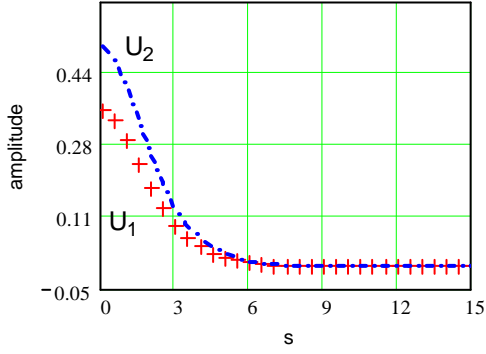

Figura 1: Resultados numéricos, em $\xi = 10$, de $U_1(s,\xi)$ (símbolo) e $U_2(s,\xi)$ (linha ponto-tracejada).

Substituindo em (7) as soluções numéricas obtidas para $U_1(s,\xi)$ e $U_2(s,\xi)$, calculamos as amplitudes complexas $a_1(s,10)$ e $a_2(s,10)$ das equações (1) e (2). Na figura 2, apresentamos as partes real e imaginária dessas amplitudes.

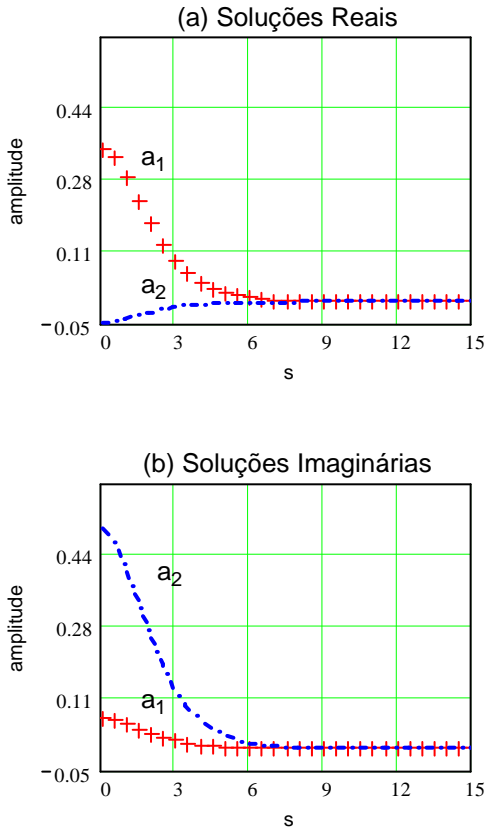

Figura 2. Soluções numéricas, em $\xi = 10$, das amplitudes $a_1(s,\xi)$ (símbolo) e $a_2(s,\xi)$ (linha ponto-tracejada). (a) Parte real das soluções, e (b) parte imaginária das soluções.

Enfim, as soluções numéricas apresentadas na figura 2 são graficamente comparadas com as soluções analíticas, conhecidas na literatura (YMAI et al., 2004), apresentadas em (5) e (6). Para realizar a comparação de forma consistente, entre as soluções numéricas e exatas, tomamos em $\xi = 10$ os valores $\alpha = -1/2$, $\beta = -1/2$, $r = -1$ e $\delta = 0$ para as quantidades dielétricas em (5) e (6). Os resultados podem ser observados nas figuras 3 e 4.

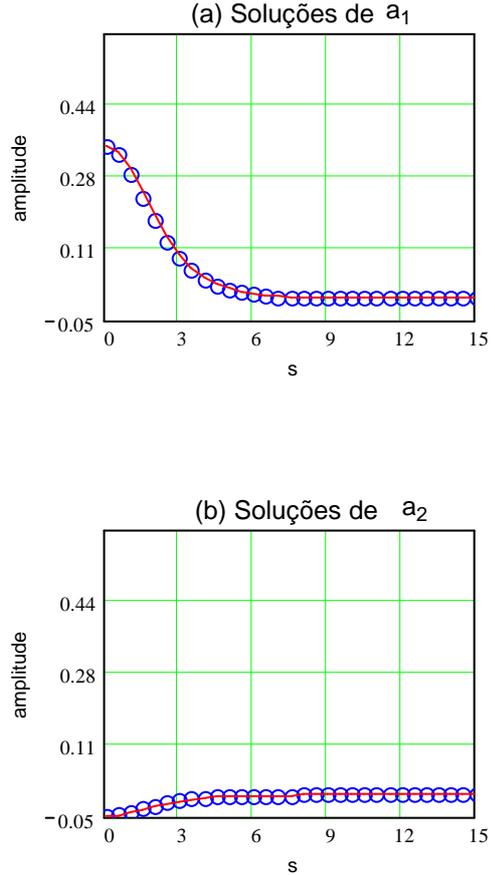

Figura 3: Comparações entre as soluções numéricas (símbolo) e analíticas (linha contínua). (a) Solução $\text{Re}(a_1)$, e (b) solução $\text{Re}(a_2)$.

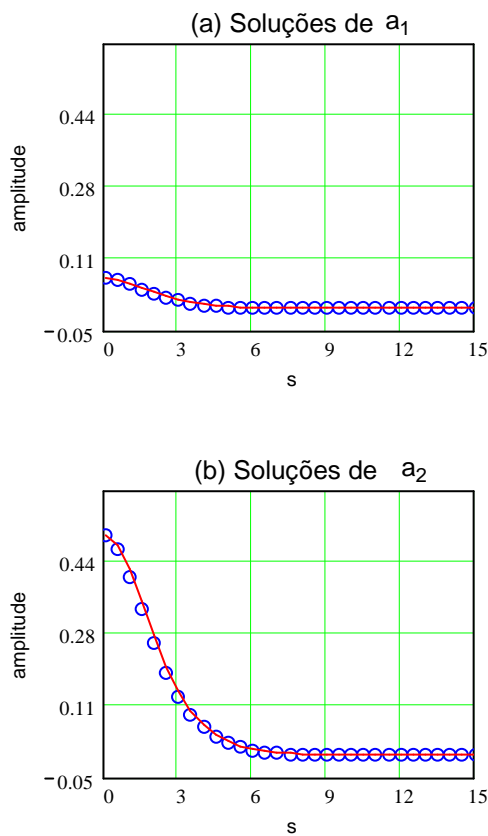

Figura 4: Comparações entre as soluções numéricas (símbolo) e analíticas (linha contínua). (a) Solução $\text{Im}(a_1)$, e (b) solução $\text{Im}(a_2)$.

## Conclusões

Verificamos que o sistema de equações diferenciais não-lineares, que descreve a evolução longitudinal de sólitons (duas ondas eletromagnéticas acopladas - modos fundamental $a_1$ e segundo harmônico $a_2$) em fibras óticas do tipo $\chi^{(2)}$, torna-se um sistema de EDP's separável, quando $\delta = 0$, cujas soluções estacionárias podem ser obtidas numericamente, veja as figuras 1 e 2, segundo o desenvolvimento numérico descrito anteriormente.

Além disto, a partir dos resultados apresentados nas figuras 3 e 4, observamos que as soluções analíticas descritas em (3) e (6), considerando $\delta = 0$, comparadas com as soluções numéricas de (13) e (14), substituídas em (7), são graficamente muito próximas. Desses resultados podemos concluir que o método numérico permitiu obter soluções aceitáveis.

Como perspectiva de continuidade deste trabalho, aplicaremos o método numérico desenvolvido na simulação da propagação de sinais do tipo sólitons ideais, ao caso de fibras óticas não-ideais, onde estudaremos como pequenas perturbações locais afetam a estabilidade destes. Sabe-se que inomogeneidades (difusão de moléculas hidrogênio, bolhas, impurezas metálicas,...) e defeitos (variação do diâmetro, rugosidade, sinuosidade no eixo, microcurvaturas, emendas, ..) da fibra ótica são as principais causas de absorção-dissipação da onda sóliton (RAGHAVAN; AGRAWAL, 2000, STROBEL, 2004). Supondo uma absorção da ordem de 2dB/Km, e um processo de bombeamento do sinal com geração de ruído, estudaremos numericamente a estabilidade da propagação de sólitons por grandes distâncias (WERNER; DRUMMOND, 1993). Nestas modelagens não-ideais a descrição analítica da propagação do sinal não é conhecida e somente um tratamento numérico poderá indicar novos caminhos para as Ciências dos Materiais Dielétricos e das Telecomunicações.

## Agradecimentos

## Referências